\documentclass[english,preprintnumbers,letterpaper,nofootinbib]{revtex4}
\usepackage[T1]{fontenc}
\usepackage[latin9]{inputenc}
\usepackage{color}
\usepackage{babel}
\usepackage{amsthm}
\usepackage{amsmath}
\usepackage{amssymb}
\usepackage{graphicx}
\usepackage[unicode=true,pdfusetitle,
 bookmarks=true,bookmarksnumbered=true,bookmarksopen=false,
 breaklinks=false,pdfborder={0 0 0},backref=false,colorlinks=true]
 {hyperref}

\makeatletter

\providecommand{\tabularnewline}{\\}

\@ifundefined{textcolor}{}
{%
 \definecolor{BLACK}{gray}{0}
 \definecolor{WHITE}{gray}{1}
 \definecolor{RED}{rgb}{1,0,0}
 \definecolor{GREEN}{rgb}{0,1,0}
 \definecolor{BLUE}{rgb}{0,0,1}
 \definecolor{CYAN}{cmyk}{1,0,0,0}
 \definecolor{MAGENTA}{cmyk}{0,1,0,0}
 \definecolor{YELLOW}{cmyk}{0,0,1,0}
 }
\theoremstyle{plain}
\newtheorem{thm}{\protect\theoremname}

\makeatother

\providecommand{\theoremname}{Theorem}

\begin{document}
\global\long\def\Tr{\mathop\mathrm{Tr}\nolimits}

\global\long\def\Span{\mathop\mathrm{span}}
\global\long\def\poly{\mathop\mathrm{poly}}

\global\long\def\cH{\mathcal{H}}

\global\long\def\kets#1{|#1\rangle}
 \global\long\def\bras#1{\langle#1|}
 \global\long\def\ket#1{\left| #1\right\rangle }
 \global\long\def\bra#1{\left\langle #1\right|}
\global\long\def\ii{\mathbb{I}}
 \global\long\def\braket#1#2{\langle#1|#2\rangle}
 \global\long\def\ep{\epsilon}

\preprint{MIT-CTP-4099}

\title{Quantum state restoration and single-copy tomography}

\author{Edward Farhi}

\email{farhi@mit.edu}

\author{David Gosset}

\email{dgosset@mit.edu}

\author{Avinatan Hassidim}

\email{avinatan@mit.edu}

\author{Andrew Lutomirski}

\email{luto@mit.edu}

\affiliation{Center for Theoretical Physics, Massachusetts Institute of Technology,
Cambridge, MA 02139}

\author{Daniel Nagaj}

\email{daniel.nagaj@savba.sk}

\affiliation{Research Center for Quantum Information, Institute of Physics, Slovak
Academy of Sciences, Dúbravská cesta 9, 845 11 Bratislava, Slovakia }

\author{Peter Shor}

\email{shor@math.mit.edu}

\affiliation{Department of Mathematics, Center for Theoretical Physics and CSAIL,
Massachusetts Institute of Technology, Cambridge, MA 02139}

\date{May 10, 2011}
\begin{abstract}
Given a single copy of an $n$ qubit quantum state $|\psi\rangle$,
the no-cloning theorem greatly limits the amount of information which
can be extracted from it. Moreover, given only a procedure which verifies
the state, for example a procedure which measures the operator $\kets{\psi}\bras{\psi}$,
we cannot prepare $|\psi\rangle$ in time polynomial in $n$. In this
paper, we consider the scenario in which we are given both a single
copy of $|\psi\rangle$ and the ability to verify it. We show that
in this setting, we can do several novel things efficiently. We present
a new algorithm that we call quantum state restoration which allows
us to extend a large subsystem of $|\psi\rangle$ to the full state,
and in turn this allows us to copy small subsystems of $|\psi\rangle$.
In addition, we present algorithms that can perform tomography on
small subsystems of $|\psi\rangle$, and we show how to use these
algorithms to estimate the statistics of any efficiently implementable
POVM acting on $|\psi\rangle$ in time polynomial in the number of
outcomes of the POVM.
\end{abstract}
\maketitle

\section{Introduction}

Quantum mechanics places constraints on what can be done with only
a single copy of an unknown state. The no-cloning theorem says that
it is impossible to copy such a state. Measuring an observable on
an unknown state generically damages it. Learning the full description
of a state or even the description of a small piece of it cannot be
done with only a single copy of it.

We are interested in the additional power given by the ability to
verify a state. Given a single copy of an unknown quantum state $\ket{\psi}$
and a verifier, that is a black box (or quantum circuit) which measures
the operator $P=\kets{\psi}\bras{\psi}$, the no-cloning theorem no
longer applies. In this setting, we present novel algorithms that
can copy small parts of the state and make measurements on $|\psi\rangle$
without damaging the state. One situation where such a verifier exists
is when $|\psi\rangle$ is the unique ground state of a particular
gapped local Hamiltonian which we know. Measuring the energy of $|\psi\rangle$
gives the ground state energy $E_{0}$. We can then use $E_{0}$ and
the Hamiltonian $H$ to verify whether any state has energy $E_{0}$.%
\footnote{Verification is not the same as measuring the energy. One way to verify
the state is to apply phase estimation, compute an indicator of whether
the energy has the right value, uncompute the phase estimation step,
and measure the indicator.%
}

To understand quantum state restoration, first consider a classical
problem. Suppose that there is some unknown $n$-bit string $z=z_{A}z_{B}$,
where $z_{A}$ is the first $n-k$ bits of $z$ and $z_{B}$ is the
last $k$ bits. Suppose further that there is a function
\[
f\left(x\right)=\begin{cases}
1 & \mbox{if }x=z\\
0 & \mbox{otherwise}
\end{cases}
\]
 on $n$-bit strings that tests whether they are equal to $z$. If
we are given $z_{A}$ and the ability to evaluate $f$, we can find
$z$ by randomly guessing: we pick a random $k$-bit string $x_{B}$
and evaluate $f\left(z_{A}x_{B}\right)$, repeating until we get $f=1$.
This finds $z$ in expected time $2^{k}$.

Quantum state restoration is a straightforward quantum generalization
of this classical algorithm, which surprisingly works even on entangled
states. If $\ket{\psi}$ lives in the Hilbert space $\mathcal{H}_{A}\otimes\mathcal{H}_{B}$,
our algorithm takes as input the part of $|\psi\rangle$ that lives
in subsystem $A$ and uses $P$ to produce as output the state $\ket{\psi}$
in expected time $O\left(\poly(\dim\mathcal{H}_{B})\right)$. It works
by randomly guessing the part of $|\psi\rangle$ that lives in subsystem
$B$ and measuring $P$. On a successful iteration (i.e.\ if the
measurement outcome is 1), then $|\psi\rangle$ is recovered. On a
failed iteration, there is minimal damage to the part of the state
in subsystem $A$ and we can try again.

This can be used to copy small subsystems of $\ket{\psi}$: if $|\psi\rangle$
has the reduced density matrix $\rho_{B}$ on a small subsystem $B$,
we can set aside subsystem $B$ and then use state restoration to
extend subsystem $A$ to the full state $|\psi\rangle$. We are left
with $|\psi\rangle$ and a mixed state $\rho_{B}$. If we use this
to obtain multiple copies of $\rho_{B}$, we can perform tomography
on subsystem $B$. We call this application single-copy tomography,
and we give two more specialized algorithms to do the same thing.
All these algorithms have running time polynomial in the dimension
of subsystem $B$. We also give a reduction from estimating the statistics
of a general POVM measurement (even if it includes noncommuting operators)
to single-copy tomography, with running time polynomial in the number
of POVM operators.

Our original motivation for developing these algorithms was to understand
the security of a class of public-key quantum money schemes. Public-key
quantum money is a quantum state that can be produced by a bank and
verified by anyone---ideally, the verification algorithm is a projector
onto the state in question \cite{QMON09,aaronson-quantum-money,mosca-2009}.
The definition of quantum money requires that no one other than the
bank can efficiently produce states that pass verification, and when
a state passes verification it is returned undamaged by the procedure.
Whether or not secure quantum money protocols exist is an open question.
However, algorithms such as quantum state restoration and single-copy
tomography rule out a large class of possible quantum money schemes.

The simplest example of a quantum money scheme that is broken by our
algorithm is based on product states. The bank chooses a string of
$n$ uniformly random angles $\theta_{i}$ between $0$ and $2\pi$.
This string is a classical secret known only to the bank. Using these
angles, the bank generates the state $|\psi\rangle=\otimes_{i}|\theta_{i}\rangle$
where $|\theta_{i}\rangle=\cos\theta_{i}|0\rangle+\sin\theta_{i}|1\rangle$
and chooses a set of (say) 4-local projectors $\left\{ P_{i}\right\} $
which are all orthogonal to $|\psi\rangle$. This set is chosen to
be large enough so that $\ket{\psi}$ is the only state in the intersection
of the zero eigenspaces of all of the projectors. The quantum money
consists of the state $|\psi\rangle$ and a classical description
of the projectors%
\footnote{The bank must also digitally sign the description of the projectors
using a classical digital signature protocol which is a secure against
quantum adversaries. Such protocols are believed to exist.%
}. The bank must choose a new set of angles $\left\{ \theta_{i}\right\} $
for each quantum money state it produces; otherwise standard tomography
can break this protocol. Anyone can verify the money by measuring
the projectors. Since a good money state is an eigenstate of the projectors,
the measurement passes along good money undamaged.

At first glance, product state quantum money seems promising. First,
given only the state $\ket{\psi}$, the no-cloning theorem prevents
anyone from making a second copy. In general, given only a set of
4-local projectors, the problem of finding the corresponding angles
(if they exist) is NP-complete (although in our case the projectors
are chosen from a specific distribution and there is a planted solution,
so the problem may be easier). However, given both the state $\ket{\psi}$
and the projectors, $\ket{\psi}$ can be efficiently copied using
quantum state restoration. We use the quantum money's verifier as
our projector $P=\kets{\psi}\bras{\psi}$. We can then copy the qubits
one at a time. To copy the first qubit, a simplified version of quantum
state restoration proceeds as follows: 
\begin{enumerate}
\item Set aside the first qubit. We are left with the state $\ket{\theta_{2}}\cdots\ket{\theta_{n}}$. 
\item Add a new register at the beginning containing a random one-qubit
state. We now have a state which can be written as 
\[
\left(\alpha\ket{\theta_{1}}+\beta\ket{\theta_{1}^{\perp}}\right)\ket{\theta_{2}}\cdots\ket{\theta_{n}}
\]
 where $\braket{\theta_{i}}{\theta_{i}^{\perp}}=0$ and $\alpha$
and $\beta$ are unknown random variables. 
\item Verify the quantum money. This produces either the desired state $|\psi\rangle$
or an invalid quantum money state 
\[
\ket{\theta_{1}^{\perp}}\ket{\theta_{2}}\cdots\ket{\theta_{n}}
\]
 with equal probability (averaged over the choice of the random state
in step 2). If we have produced the desired state, then we have cloned
the first qubit: we have both the copy in the $|\psi\rangle$ and
the qubit that we set aside in step 1. If not, then we discard the
qubit $|\theta_{1}^{\perp}\rangle$ and go back to step two and repeat
until we get $|\theta_{1}\rangle$.
\end{enumerate}
Repeating this procedure for each qubit allows us to clone the state
$\ket{\psi}$ in linear time.

The algorithm we just described can copy the full $n$-qubit state
$\ket{\psi}$ because $\ket{\psi}$ is a product state. We can think
of this algorithm as first removing a subsystem of $\ket{\psi}$ (step
1) and then recovering the state $\ket{\psi}$ from the part that
remains (steps 2 and 3). Surprisingly, a small modification of steps
2 and 3 leads to an algorithm that efficiently restores small missing
subsystems, even on entangled states: this is quantum state restoration.

Our paper is structured as follows. In section~\ref{sec:Quantum-state-restoration},
we present the quantum state restoration algorithm and analyze its
running time. In section~\ref{sec:Single-copy-tomography}, we present
two alternative algorithms for single-copy tomography, one of which
is asymptotically faster than quantum state restoration. Finally,
we give several scenarios in which new algorithms could be developed
using the techniques in this paper.

\section{Quantum state restoration\label{sec:Quantum-state-restoration}}

Quantum state restoration takes as input a large subsystem of a state
$|\psi\rangle$ (this subsystem could be, for example, the first $n-k$
qubits of the $n$ qubit state $|\psi\rangle)$ and, using the ability
to measure the projector $P=|\psi\rangle\langle\psi|$, reconstructs
the full state $|\psi\rangle$.
\begin{thm}
\label{thm:state-restoration}Suppose that $|\psi\rangle$ is an unknown
quantum state in a Hilbert space $\mathcal{H}_{A}\otimes\mathcal{H}_{B}$
and we are given oracle access to a coherent measurement of $P=|\psi\rangle\langle\psi|$
(that is, the oracle performs the operation $P\otimes\ii+\left(1-P\right)\otimes\sigma_{x}$
on the original Hilbert space plus a single-qubit ancilla). Then there
exists an efficient quantum algorithm that takes as input a mixed
state in $\mathcal{H}_{A}$ with density matrix $\Tr_{B}|\psi\rangle\langle\psi|$
and outputs $|\psi\rangle$. This algorithm makes an expected number
$O\left(\left(\dim\mathcal{H}_{B}\right)^{2}\right)$ of calls to
the measurement oracle.
\end{thm}
The idea is that any state $|\psi\rangle$ on a Hilbert space $\mathcal{H}_{A}\otimes\mathcal{H}_{B}$
(where $d$ is the dimension of $\mathcal{H}_{B})$ can be Schmidt
decomposed as
\[
|\psi\rangle=\sum_{i=1}^{\chi}\sqrt{p_{i}}|u_{i}\rangle|v_{i}\rangle
\]
 where $\chi$ is the Schmidt rank of $|\psi\rangle$ (note that $\chi\le d$).
If we start with the state $|\psi\rangle$ and set aside the part
that lives on $\mathcal{H}_{B}$, then we are left with the mixed
state $\rho_{A}=\Tr_{B}|\psi\rangle\langle\psi|$, which has all of
its support on the Schmidt basis $\Span\left\{ |u_{i}\rangle\right\} $.
From $\rho_{A}$, we can construct the state $\rho_{A}\otimes\frac{I}{d}$
on $\mathcal{H}_{A}\otimes\mathcal{H}_{B}$. We now measure the projector
$P$. If we obtain the outcome 1, then we are left with the state
$|\psi\rangle,$ If not, we discard (i.e. trace out) $\mathcal{H}_{B}$,
leaving a state on $\mathcal{H}_{A}$ that \emph{still} has all of
its support on the Schmidt basis. We then try again until we obtain
the outcome 1. If all the $p_{i}$ are equal, then each attempt succeeds
with probability $\frac{1}{\chi d}$, and the entire algorithm finishes
in an expected number of iterations $\chi d$. For general values
$\left\{ p_{i}\right\} $, the expected running time is still exactly
$\chi d$, although the distribution of the running time becomes more
complicated.

We now summarize the quantum state restoration algorithm. 
\begin{enumerate}
\item Start with the state $|\psi\rangle\in\mathcal{H}_{A}\otimes\mathcal{H}_{B}$
and set aside the part of $|\psi\rangle$ that lives in subsystem
$B$. We are left with the mixed state 
\[
\rho_{A}=\Tr_{B}|\psi\rangle\langle\psi|.
\]

\item Add a random state on subsystem $B$. The state is now 
\[
\rho_{A}\otimes\frac{\ii}{d}.
\]
 
\item Measure the projector $P=|\psi\rangle\langle\psi|.$ If the outcome
is $+1$ then you are done: you still have the original copy of subsystem
$B$ that you set aside and you have recovered the state $|\psi\rangle.$
If not, discard subsystem $B$ and repeat from step 2.
\end{enumerate}
We now show that the expected running time of this algorithm is $\chi\cdot d\le d^{2}$
(measured in number of uses of $P$).

\subsection{Running Time of Quantum State Restoration\label{sub:Expected-Running-Time}}

\begin{figure}[t]
\centering{}\includegraphics[scale=0.75]{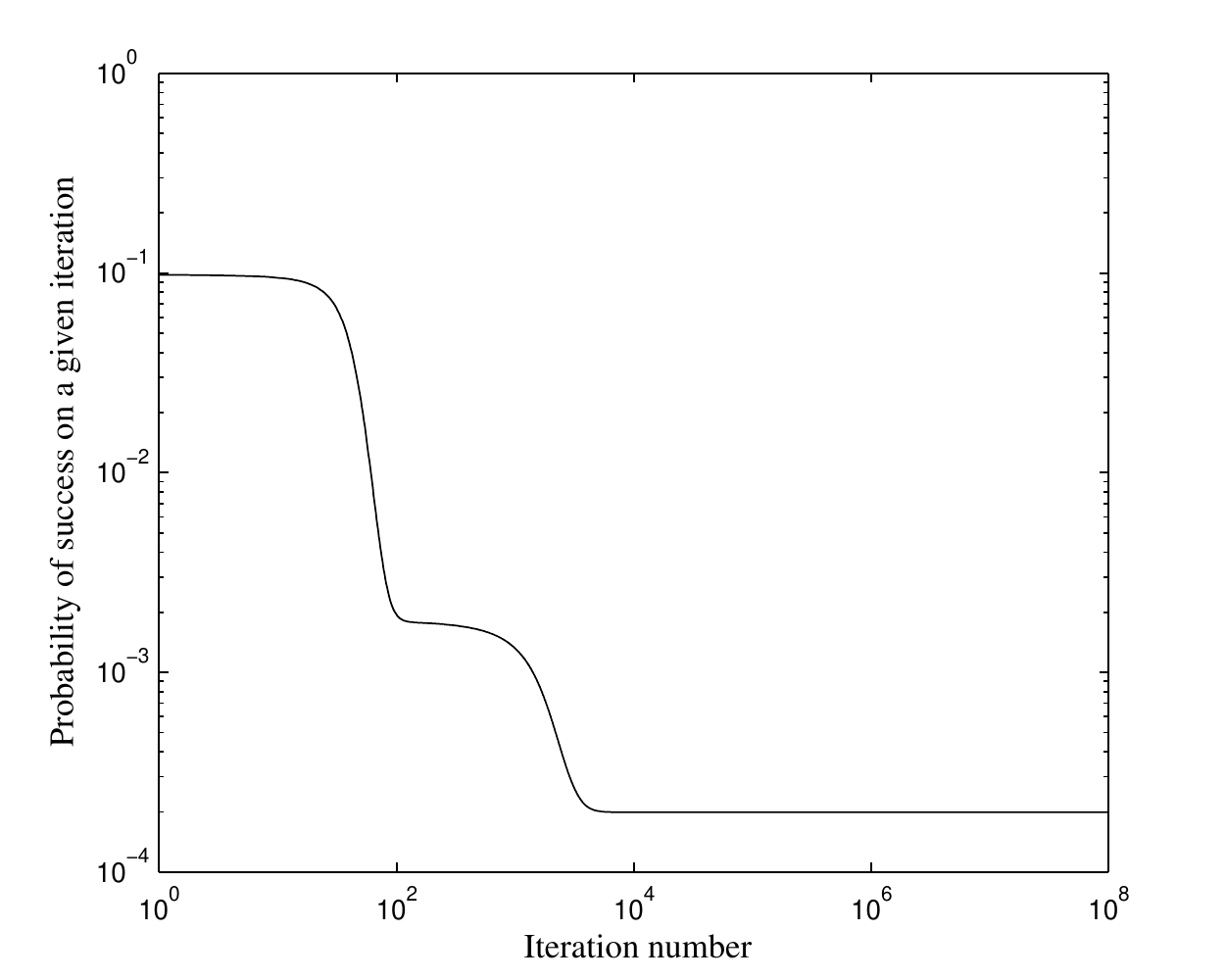}\caption{Probability of restoring the state on a given iteration conditioned
on all previous iterations failing. Conditioned on failing every time,
the first two flat regions are metastable states and the third is
stable. In this graph, $|\psi\rangle=\sqrt{1-10^{-2}-10^{-4}}|0\rangle_{A}|0\rangle_{B}+\sqrt{10^{-2}}|1\rangle_{A}|1\rangle_{B}+\sqrt{10^{-4}}|2\rangle_{A}|2\rangle_{B},$
$\dim\mathcal{H}_{B}=10$, and the expected number of iterations required
is 30.\label{fig:running-time}}
\end{figure}

In the simple case where all of the $p_{i}$ are equal, then the initial
state $\rho_{A}$ is the fully mixed state over the $\Span\left\{ |u_{i}\rangle\right\} $.
In this case, if you measure 0 in step 3, the density matrix left
in register $A$ after discarding register $B$ is unchanged. The
algorithm terminates with probability $\frac{1}{\chi\cdot d}$ on
each iteration, finishing in an expected number of iterations $\chi\cdot d$.
If the $p_{i}$ are not all equal, then the algorithm can reach bad
states where most of the weight is on low-weight elements of the Schmidt
basis. When this happens, the chance of success on any given iteration
drops (see Fig.~\ref{fig:running-time} for an extreme example),
but the probability of reaching these bad states decreases with the
corresponding $p_{i}$. Surprisingly, these effects exactly cancel,
and the expected number of iterations required to restore the state
is $\chi\cdot d$ regardless of the values of the $p_{i}$.

To prove this, we define two maps 
\begin{align*}
F_{0}(\sigma) & =\Tr_{B}\left[\left(1-|\psi\rangle\langle\psi|\right)\left(\sigma\otimes\frac{\ii}{d}\right)\left(1-|\psi\rangle\langle\psi|\right)\right]\\
F_{1}(\sigma) & =\Tr_{B}\left[|\psi\rangle\langle\psi|\left(\sigma\otimes\frac{\ii}{d}\right)|\psi\rangle\langle\psi|\right].
\end{align*}
 Here $F_{b}(\sigma$) is the unnormalized density matrix obtained
by measuring $P$ on the state given by the density matrix $\sigma$,
conditioned on the measurement outcome $b\in\left\{ 0,1\right\} $.
The probability of obtaining a sequence of measurement outcomes $b_{1},b_{2},\ldots,b_{m}$,
starting with the state $\sigma$ is then given by 
\begin{equation}
\Pr\left[\{b_{1,}b_{2},b_{3},...,b_{m}\}\,\big|\,\sigma\right]=\Tr[F_{b_{m}}\circ\cdots\circ F_{b_{1}}(\sigma)],\label{eq:pr_k}
\end{equation}
 which can be seen by induction: 
\begin{align*}
\Pr\left[\{b_{1,}b_{2},b_{3},...,b_{m}\}\,\big|\,\sigma\right] & =\Pr\left[b_{m}\,\big|\,\sigma,\{b_{1,}b_{2},b_{3},...,b_{m-1}\}\right]\\
 & \quad\times\Pr\left[\{b_{1,}b_{2},b_{3},...,b_{m-1}\}\,\big|\,\sigma\right]\\
 & =\Tr F_{b_{m}}\left(\frac{F_{b_{m-1}}\circ\cdots\circ F_{b_{1}}\left(\sigma\right)}{\Tr\left(F_{b_{m-1}}\circ\cdots\circ F_{b_{1}}\left(\sigma\right)\right)}\right)\\
 & \quad\times\Tr\left(F_{b_{m-1}}\circ\cdots\circ F_{b_{1}}\left(\sigma\right)\right)\\
 & =\Tr F_{b_{m}}\left(F_{b_{m-1}}\circ\cdots\circ F_{b_{1}}\left(\sigma\right)\right).
\end{align*}
 We can use this equation to write an explicit formula for the expected
number of measurements $T\left(\sigma\right)$, starting with the
state $\sigma$:
\begin{align}
T(\sigma) & =\sum_{k=1}^{\infty}k\cdot\Pr\big[\{\underset{k-1}{\underbrace{0,0,\ldots,0}},1\}\,\big|\,\sigma\big]\nonumber \\
 & =\sum_{k=1}^{\infty}k\cdot\Tr\left[F_{1}\circ F_{0}\circ\cdots\circ F_{0}(\sigma)\right].\label{eq:explicit-E}
\end{align}
 As written, this formula is difficult to evaluate, but we can see
that it is linear in $\sigma$. We are interested in the quantity
$T(\rho_{A})$, which we expand as 
\begin{eqnarray}
T(\rho_{A})=\sum_{i=1}^{\chi}p_{i}T(|u_{i}\rangle\langle u_{i}|).\label{eq:linearity}
\end{eqnarray}
 We expand $T(|u_{i}\rangle\langle u_{i}|)$ by conditioning on the
outcome of the first measurement: 
\begin{eqnarray*}
T\left(|u_{i}\rangle\langle u_{i}|\right) & = & \Pr\left[1\,\big|\,|u_{i}\rangle\langle u_{i}|\right]+\Pr\left[0\,\big|\,|u_{i}\rangle\langle u_{i}|\right]\left(1+T\left(\frac{F_{0}(|u_{i}\rangle\langle u_{i}|)}{\Pr\left[0\,\big|\,|u_{i}\rangle\langle u_{i}|\right]}\right)\right)\\
 & = & 1+T\left(F_{0}(|u_{i}\rangle\langle u_{i}|)\right)\\
 & = & 1+T\left(|u_{i}\rangle\langle u_{i}|-2\frac{p_{i}}{d}|u_{i}\rangle\langle u_{i}|+\frac{p_{i}}{d}\sum_{j=1}^{\chi}p_{j}|u_{j}\rangle\langle u_{j}|\right)\\
 & = & 1+\left(1-2\frac{p_{i}}{d}\right)T\left(|u_{i}\rangle\langle u_{i}|\right)+\frac{p_{i}}{d}\sum_{j=1}^{\chi}p_{j}T\left(|u_{j}\rangle\langle u_{j}|\right).
\end{eqnarray*}
 Using \eqref{eq:linearity}, this can be transformed into 
\[
2p_{i}T\left(|u_{i}\rangle\langle u_{i}|\right)-p_{i}T(\rho_{A})=d.
\]
 Summing both sides over $i=1,\dots,\chi$ using $\sum p_{i}=1$ and
\eqref{eq:linearity} again, we obtain 
\[
T(\rho_{A})=\chi\cdot d,
\]
 which is the desired result. This proves theorem~\ref{thm:state-restoration}.

\section{Single-copy tomography and estimation of measurement statistics\label{sec:Single-copy-tomography}}

We expect that quantum state restoration will most commonly be used
to perform tomography on a single copy of a verifiable quantum state.
We can perform several different types of tomography, and we give
algorithms for some types that are faster than quantum state restoration.

\vspace{0.7em}

\noindent \textbf{General tomography on a subsystem}

\vspace{0.7em}

In the simplest case, we have a single copy of an unknown state $\ket{\psi}$
and access to the measurement $P=\kets{\psi}\bras{\psi}$ and we would
like to estimate properties of the density matrix $\rho_{B}=\Tr_{A}\kets{\psi}\bras{\psi}$
for a subsystem $B$. We can do this by using quantum state restoration
to prepare many unentangled states, each with (independent) density
matrixes $\rho_{B}$. We can then use any standard state tomography
algorithm on these states.

\vspace{0.7em}

\noindent \textbf{Measurement of a subsystem in an orthogonal basis}

\vspace{0.7em}

For many applications, it is sufficient to estimate the probabilities
$q_{i}=\Tr\left[|i\rangle_{B}\langle i|_{B}|\psi\rangle\langle\psi|\right]$
of obtaining the outcome $i$ if one were to measure subsystem $B$
of $|\psi\rangle$ in the orthonormal basis $\left\{ |i\rangle_{B}\right\} $.
Quantum state restoration can sample these probabilities directly.
We discuss this application in section~\ref{sub:tomograph-with-state-restoration}.

In sections \ref{sub:tomograph-with-state-restoration} and \ref{sub:improved-algos},
we present two other specialized algorithms to compute these probabilities.
Both algorithms measure the $q_{i}$ one at a time by considering
the statistics of the two-outcome measurements $\left\{ |i\rangle_{B}\langle i|_{B},\ii-|i\rangle_{B}\langle i|_{B}\right\} $,
and both are based on previously presented schemes for amplifying
QMA verifiers \cite{MWqma,NWZqma}.

In each case, we fix a precision $\delta>0$ and an error probability
$\epsilon>0$ and compute the running time to produce estimates $q_{i}^{\text{est}}$
such that 
\[
|q_{i}^{\text{est}}-q_{i}|<\delta
\]
 for all $i$ with probability at least $1-\epsilon$.

\vspace{0.7em}

\noindent \textbf{Estimation of the statistics of any POVM}

\vspace{0.7em}

We can use any of our algorithms to estimate the statistics of a general
measurement (on the complete state, not just a subsystem). This is
because a general POVM measurement can be reduced to a measurement
of a subsystem in an orthogonal basis, as we now review. Given an
efficiently implementable POVM $\left\{ E_{i}\right\} $ where $i\in\left\{ 1,\ldots,d\right\} $,
we can implement a unitary operator $U$ such that 
\[
U\left(|\phi\rangle_{A}|1\rangle_{B}\right)=\sum_{i=1}^{d}\left(\sqrt{E_{i}}|\phi\rangle_{A}\right)|i\rangle_{B}
\]
 for any state $|\phi\rangle$. If we work in a two-register Hilbert
space, where register $A$ can hold $|\phi\rangle$ and register $B$
has dimension $d$, then the probability of measurement outcome $i$
when the POVM is measured on $|\phi\rangle$ is equal to 
\begin{align*}
\langle\phi|E_{i}|\phi\rangle & =\Tr\left[\rho_{B}\kets i_{B}\bras i_{B}\right]
\end{align*}
 where $\rho_{B}=\Tr_{A}\left[U|\phi\rangle_{A}|1\rangle_{B}\langle1|_{B}\langle\phi|_{A}U^{\dagger}\right]$.
If we define 
\begin{align*}
|\psi\rangle & =U|\phi\rangle_{A}|1\rangle_{B}\\
P' & =|\psi\rangle\langle\psi|=UPU^{\dagger}
\end{align*}
 then $|\psi\rangle$ can be efficiently prepared (given $|\phi\rangle$)
and $P'$ can be efficiently measured. Now we can use any of the algorithms
to estimate the measurement statistics of subsystem $B$ of $|\psi\rangle$
using the projector $P'$ in the computational basis (that is, any
of the algorithms below) to estimate the probabilities $\langle\phi|E_{i}|\phi\rangle=\Tr\left[\kets i\bras i\rho'_{B}\right]$.
After estimating the probabilities, we uncompute $U$ to recover the
initial state $|\phi\rangle$. We summarize this ability with the
following theorem.
\begin{thm}
\label{thm:single-copy-tomography}Suppose that $|\phi\rangle$ is
an unknown quantum state and we are given oracle access to a coherent
measurement of $P=|\phi\rangle\langle\phi|$ (that is, the oracle
performs the operation $P\otimes\ii+\left(1-P\right)\otimes\sigma_{x}$
on the original Hilbert space plus a single-qubit ancilla). Fix $0<\epsilon<1$,
$\delta>0$, and an efficiently implementable $d$-outcome POVM given
by operators $\left\{ E_{i}\right\} $. Then there exists an efficient
quantum algorithm that takes as input a single copy of $|\phi\rangle$
and outputs an undamaged copy of $|\phi\rangle$ along with estimates
$q_{i}^{\text{est}}$ such that 
\[
|q_{i}^{\text{est}}-\langle\phi|E_{i}|\phi\rangle|<\delta
\]
 for all $i$ with probability at least $1-\epsilon$. This algorithm
uses an expected number $O\left(\frac{d}{\delta}\log\left(\frac{d}{\epsilon}\right)\right)$
calls to the measurement oracle and the POVM.
\end{thm}
The algorithm which achieves this running time is given in section~\ref{sub:phase-est-algo}.

If we want to perform tomography on a subsystem of $|\phi\rangle$,
we can use theorem~\ref{thm:single-copy-tomography} to estimate
an informationally complete POVM on that subsystem.

\subsection{Using quantum state restoration to estimate measurement statistics
\label{sub:tomograph-with-state-restoration}}

In this section we consider the running time of estimating the probabilities
$q_{i}=\Tr\left[\rho_{B}|i\rangle_{B}\langle i|_{B}\right]$ on a
given state $|\psi\rangle$ using quantum state restoration. We do
this by repeatedly measuring register $B$ and then restoring the
state. Let $m_{i}$ be the number of times we observe outcome $i$
in $N$ trials. Our estimate of $q_{i}$ is 
\[
q_{i}^{\text{est}}=\frac{m_{i}}{N}.
\]

For the $j^{\text{th}}$ observation, let $x_{i,j}\in\left\{ 0,1\right\} $
indicate whether the outcome of that observation was $i$. For fixed
$i$, the $x_{i,j}$ are independent. To obtain a bound on the error
$|q_{i}^{\text{est}}-q_{i}|$, we use Hoeffding's inequality \cite{Hoeffding},
which for a sequence of $N$ independent and identically distributed
random bits $x_{i,j}$ with mean value $\mathop\mathbb{E}\nolimits _{j}[x_{i,j}]=q_{i}$
implies that 
\begin{equation}
\Pr\left[\left|\frac{1}{N}\sum_{j=1}^{N}x_{i,j}-q_{i}\right|\geq\delta\right]\leq2e^{-2N\delta^{2}}\text{, for any }\delta>0.\label{eq:hoeffding}
\end{equation}
 So 
\[
\Pr\left[|q_{i}^{\text{est}}-q_{i}|\geq\delta\right]\leq2e^{-2N\delta^{2}}
\]
 for each $i$ individually, and, by a union bound, 
\[
\Pr\left[|q_{i}^{\text{est}}-q_{i}|\geq\delta\text{ for any }i\right]\leq2de^{-2N\delta^{2}}.
\]
 Choosing $N=\left\lceil \frac{1}{2\delta^{2}}\ln\frac{2d}{\epsilon}\right\rceil $
makes the right hand side $\leq\epsilon$. Each of the $N$ repetitions
of quantum state restoration takes an expected time $\chi\cdot d$,
so the total expected number $\mathbb{E}[M_{\text{SR}}]$ (where the
subscript stands for {}``state restoration'') of uses of $P$ is
\[
\mathbb{E}[M_{\text{SR}}]=\chi\cdot d\left\lceil \frac{1}{2\delta^{2}}\ln\frac{2d}{\epsilon}\right\rceil .
\]

\subsection{Improved algorithms to estimate measurement statistics \label{sub:improved-algos}}

In this section we describe two other algorithms which can be used
for single-copy tomography. Both of these approaches are based on
Jordan's lemma \cite{JordansLemma}. The algorithms we discuss in
this section are based on the QMA amplification schemes of Marriott
and Watrous \cite{MWqma} and Nagaj et al.\ \cite{NWZqma}.

To use these algorithms, we fix $i\in\left\{ 1,\ldots,d\right\} $
and we will estimate 
\[
q_{i}=\Tr\left[\rho_{B}|i\rangle_{B}\langle i|_{B}\right].
\]
 We repeat this for each value of $i$.

We begin by defining the projector 
\[
Q_{i}=|i\rangle_{B}\langle i|_{B}
\]
 and the states 
\begin{eqnarray*}
|v_{i}\rangle & = & \frac{1}{\sqrt{q_{i}}}Q_{i}|\psi\rangle,\\
|v_{i}^{\perp}\rangle & = & \frac{1}{\sqrt{1-q_{i}}}\left(1-Q_{i}\right)|\psi\rangle.
\end{eqnarray*}
Note that we can write 
\begin{eqnarray}
|\psi\rangle & = & \sqrt{q_{i}}|v_{i}\rangle+\sqrt{1-q_{i}}|v_{i}^{\perp}\rangle.\label{eq:psi}
\end{eqnarray}
We also define the state 
\begin{equation}
|\psi_{i}^{\perp}\rangle=-\sqrt{1-q_{i}}|v_{i}\rangle+\sqrt{q_{i}}|v_{i}^{\perp}\rangle.\label{eq:psi_perp}
\end{equation}
We can then use the above expressions to write $|v_{i}\rangle$ and
$|v_{i}^{\perp}\rangle$ in terms of $|\psi\rangle$ and $|\psi_{i}^{\perp}\rangle$
\begin{eqnarray}
|v_{i}\rangle & = & \sqrt{q_{i}}|\psi\rangle-\sqrt{1-q_{i}}|\psi_{i}^{\perp}\rangle\nonumber \\
|v_{i}^{\perp}\rangle & = & \sqrt{1-q_{i}}|\psi\rangle+\sqrt{q_{i}}|\psi_{i}^{\perp}\rangle.\label{eq:v_and_vperp}
\end{eqnarray}
 The principal angle $\theta_{i}\in[0,\frac{\pi}{2}]$ between the
two bases $\{|\psi\rangle,|\psi_{i}^{\perp}\rangle\}$ and $\{|v_{i}\rangle,|v_{i}^{\perp}\rangle\}$
is defined by 
\begin{equation}
\cos^{2}\theta_{i}=\left|\braket{v_{i}}{\psi}\right|^{2}=\langle\psi|v_{i}\rangle\langle v_{i}|\psi\rangle=\bra{\psi}Q_{i}\ket{\psi}=q_{i}.\label{eq:defT}
\end{equation}
 Having defined the two bases $\{|v_{i}\rangle,|v_{i}^{\perp}\rangle\}$
and $\{|\psi\rangle,|\psi_{i}^{\perp}\rangle\}$, we are now ready
to describe two algorithms for computing the expectation value $q_{i}$
more efficiently than by using quantum state restoration. For any
chosen $\epsilon$ and $\delta$, each of these algorithms will generate
an estimate $q_{i}^{\text{est}}$ such that $|q_{i}^{\text{est}}-q_{i}|<\delta$
with probability at least $1-\frac{\epsilon}{d}$. Repeating for each
$i$, we have $|q_{i}^{\text{est}}-q_{i}|<\delta$ for all $i$ with
probability at least $1-\epsilon$ by a union bound. The running times
of these algorithms as a function of $\delta$ and $\epsilon$ are
summarized in Table~\ref{tab:Comparison-of-algorithms}.

\subsubsection{Alternating Projections}

This algorithm is an application of the scheme of Marriott and Watrous
\cite{MWqma} which was originally proposed for witness-reusing amplification
of the complexity class QMA. Observe from \eqref{eq:psi}, \eqref{eq:psi_perp}
and \eqref{eq:v_and_vperp} that when performing the measurement $P$
on the state $\ket{v_{i}}$, the probability of obtaining 1 (and the
state $\ket{\psi}$) is $q_{i}$. Similarly, when measuring $P$ on
the state $\ket{v_{i}^{\perp}}$, the probability of obtaining 0 (and
the state $\ket{\psi_{i}^{\perp}}$) is also $q_{i}$. We can estimate
$q_{i}$ by performing many alternating measurements of $P$ and $Q_{i}$
and counting the number of transitions $\ket{v_{i}}\leftrightarrow\ket{\psi}$
or $|v_{i}^{\perp}\rangle\leftrightarrow|\psi_{i}^{\perp}\rangle$.
Let us now present the algorithm and compute its complexity measured
by the expected number of measurements of $P$, as a function of the
desired precision $\delta$ and error probability $\frac{\epsilon}{d}$.
\begin{enumerate}
\item Start with the state $|\psi\rangle$. Fix $N=\left\lceil \frac{1}{2}+\frac{\ln\frac{2d}{\epsilon}}{4\delta^{2}}\right\rceil $.
\item Repeat for $t=1,\ldots,N$

\begin{enumerate}
\item Measure $Q_{i}$ and record the measurement outcome as a bit $a_{2t-1}\in\{0,1\}$.
This produces one of the two states $|v_{i}\rangle$ or $|v_{i}^{\perp}\rangle$.
\item Measure the projector $P=|\psi\rangle\langle\psi|$ and record the
result $a_{2t}\in\{0,1\}$. This produces either the state $|\psi\rangle$
or $|\psi_{i}^{\perp}\rangle$ .
\end{enumerate}
\item If the state is not currently $|\psi\rangle$ (because the last measurement
in step 2b gave a $0$), then the state is $|\psi_{i}^{\perp}\rangle$.
In this case alternate measuring $Q_{i}$ and $P$ until you recover
$|\psi\rangle$.
\item From the list $\left(a_{1},\ldots,a_{2N}\right)$, compute the list
of differences $(\Delta_{1},\Delta_{2},...,\Delta_{2N-1})$ where
$\Delta_{j}=a_{j+1}\oplus a_{j}$. Let $m$ denote the number of zeros
in this list of differences. Then the estimate of $q$ is given by
\begin{equation}
q_{i}^{\text{est}}\equiv\frac{m}{2N-1}.\label{eq:q_est_naive}
\end{equation}
 
\end{enumerate}
As discussed above, the probability of getting a measurement outcome
($1$ or $0)$ which is the same as the previous measurement outcome
is $q_{i}$. So the number of zeros which appear in the list $(\Delta_{1},\Delta_{2},...,\Delta_{2N-1})$
is a binomial random variable with mean $q_{i}(2N-1)$. This is why
\eqref{eq:q_est_naive} gives an estimator for the value of $q_{i}$.

We now show that the estimate $q_{i}^{\text{est}}$ from \eqref{eq:q_est_naive}
has the required precision $\delta$, with probability at least $1-\epsilon.$
To show this, we again use Hoeffding's inequality \eqref{eq:hoeffding}.
Applying this to the case at hand with $q_{k}=1\oplus\Delta_{k}$
for $k\in\{1...,2N-1\}$, we obtain 
\[
\Pr\left[|q_{i}^{\text{est}}-q_{i}|\geq\delta\right]\leq2e^{-2(2N-1)\delta^{2}}.
\]
 The choice $N=\left\lceil \frac{1}{2}+\frac{\log\frac{2d}{\epsilon}}{4\delta^{2}}\right\rceil $
guarantees that the right hand side is $\leq\frac{\epsilon}{d}$.
Thus we have shown that the desired precision $\delta$ is achieved
by our scheme with probability at least $1-\frac{\epsilon}{d}$.

We now derive the expected number $\mathbb{E}[M_{\text{AP}}^{\left(i\right)}]$
(AP stands for alternating projections) of uses of $P$ in the above
algorithm. The random variable $M_{\text{AP}}^{\left(i\right)}$ is
$N$ plus the number of additional uses of $P$ in step 3. The operation
composed of measuring $Q_{i}$ and then measuring $P$ is an update
of a symmetric random walk on the two states $\left\{ \ket{\psi},|\psi_{i}^{\perp}\rangle\right\} $.
Let $w\left(r\right)$ be the probability of transitioning from $\ket{\psi}$
to $|\psi_{i}^{\perp}\rangle$ in $r$ steps. Then with probability
$1-w\left(N\right)$ step 3 does not use $P$ at all and, with probability
$w\left(N\right)$ it uses an expected number $\frac{1}{w\left(1\right)}$
invocations of $P$. Thus the expected running time of the algorithm
is 
\begin{align*}
\mathbb{E}\left[M_{\text{AP}}^{\left(i\right)}\right] & =N+w\left(N\right)\frac{1}{w\left(1\right)}\\
 & \le2N.
\end{align*}
In the last line, we used the fact that $w\left(N\right)$ is less
than or equal to the probability of at least one transition occurring
in $N$ steps, which is at most $Nw\left(1\right)$ by a union bound.

Hence 
\[
\mathbb{E}[M_{\text{AP}}^{\left(i\right)}]\leq2\left(\left\lceil \frac{1}{2}+\frac{1}{4\delta^{2}}\ln\frac{2d}{\epsilon}\right\rceil \right).
\]
 Repeating this procedure to obtain estimates of each $q_{i}$ (which
are all within the desired precision $\delta$ with probability at
least $1-\epsilon$) takes the expected running time 
\[
\mathbb{E}[M_{\text{AP}}]\leq2d\left(\left\lceil \frac{1}{2}+\frac{1}{4\delta^{2}}\ln\frac{2d}{\epsilon}\right\rceil \right).
\]

\subsubsection{An improved algorithm using phase estimation\label{sub:phase-est-algo}}

In this section we will give an improved algorithm for single-copy
tomography using phase estimation, based on a fast QMA amplification
scheme given in \cite{NWZqma}. Its advantage over the previous two
algorithms is that it requires quadratically fewer measurements of
$P$ . The results of this section will prove Theorem~\ref{thm:single-copy-tomography}.

As in the previous section, we estimate the $q_{i}$ one at a time
for $i\in\left\{ 1,\ldots,d\right\} $.

We begin by defining the unitary operator 
\[
W_{i}=(2P-\ii)(2Q_{i}-\ii),
\]
 which is a product of two reflections. Note that if we can implement
$P$ so that it coherently xors its measurement outcome into an ancilla
register (as in the assumption of theorem \ref{thm:single-copy-tomography}),
then we can implement the operator $\left(2P-\ii\right)$ by first
initializing that ancilla to $|-\rangle$ and applying the measurement.

Within the 2D subspace $S_{i}$ spanned by the vectors $|\psi\rangle$
and $|\psi_{i}^{\perp}\rangle$ \eqref{eq:v_and_vperp}, the operator
$W_{i}$ is a rotation 
\begin{eqnarray}
W_{i}\big|_{S_{i}} & = & e^{-2i\theta_{i}\sigma_{y}},
\end{eqnarray}
 where $\theta_{i}$ is the principal angle as defined in \eqref{eq:defT}
(and $\sigma_{y}$ refers to the Pauli matrix). 

We now describe how to obtain $q_{i}=\bra{\psi}Q_{i}\ket{\psi}=\cos^{2}\theta_{i}$
by running phase estimation of the operator $W_{i}$ on the state
$\ket{\psi}$. The eigenvectors of $W_{i}$ are 
\begin{equation}
\kets{\phi_{i}^{\pm}}=\frac{1}{\sqrt{2}}\left(\kets{\psi}\pm i\kets{\psi_{i}^{\perp}}\right).\label{eq:roteigs}
\end{equation}
 and correspond to eigenvalues $e^{\mp i2\pi\phi_{i}}$, where $\phi_{i}=\frac{\theta_{i}}{\pi}$
so that $0<\phi_{i}<\frac{1}{2}$. After running phase estimation
of $W_{i}$ on the input state $\ket{\psi}$, we will likely measure
a good approximation to either $\phi_{i}$ or $1-\phi_{i}$. Note
that either outcome provides a good estimate of 
\[
q_{i}=\cos^{2}(\pi\phi_{i})=\cos^{2}(\pi(1-\phi_{i})).
\]
This is the idea of the algorithm we present in this section. Our
algorithm must have a failure probability lower than that obtained
by a single use of phase estimation, and we must recover the state
$|\psi\rangle$ at the end of the algorithm.

Our algorithm begins by defining 
\begin{align}
t & =\left\lceil \log_{2}\left(\frac{3\pi}{\delta}\right)\right\rceil +2.\nonumber \\
r & =\left\lceil \frac{1}{\log_{2}\left(\frac{2}{\sqrt{3}}\right)}\log_{2}\left(\frac{d}{2\epsilon}\right)\right\rceil .\label{eq:r_value}
\end{align}
We proceed as follows:
\begin{enumerate}
\item Start in the state $|\psi\rangle|0\rangle^{\otimes t}$.
\item Repeat for $j=1,...,r$:

\begin{enumerate}
\item Reset the $t$ qubits of the second register to the state $|0\rangle^{\otimes t}$.
Perform phase estimation of the operator $W_{i}$ on the state of
the first register, computing the phase using the $t$ ancillae in
the second register. Define
\[
q_{i}^{(j)}=\cos^{2}\left(\pi\phi_{i}^{(j)}\right)
\]
 where $\phi_{i}^{(j)}$ is the measured phase.
\item Measure the projector $P=|\psi\rangle\langle\psi|$ on the first register. 
\end{enumerate}
\item If the state is not currently $|\psi\rangle$ (because the last measurement
in step 2(b) gave a $0$), then the state is $|\psi_{i}^{\perp}\rangle$.
In this case repeat phase estimation followed by measurement of $P$
until you measure a $1$ for $P$, recovering the state $|\psi\rangle$. 
\item Let $q_{i}^{est}$ be the median of the values $\{q_{i}^{(j)}\}$
for $j\in\{1,...r\}$.
\end{enumerate}
We now determine the expected runtime of this algorithm, and then
we will show that the resulting estimate $q_{i}^{est}$ achieves the
desired precision with high enough probability. Our analysis of the
runtime is based on the observation that each iteration of phase estimation
followed by measurement of $P$ is an update of a random walk on the
two states $\{|\psi\rangle,|\psi_{i}^{\perp}\rangle\}.$ If we start
in state $|\psi\rangle$ of the first register then after applying
phase estimation (but before measuring the phase) we obtain a state
\[
|\Psi_{i}\rangle=\frac{1}{\sqrt{2}}\left(|\phi_{i}^{+}\rangle|\gamma\rangle+|\phi_{i}^{-}\rangle|\mu\rangle\right).
\]
 where $|\gamma\rangle$ and $|\mu\rangle$ are $t$-qubit states.
So the probability of measuring 1 in step 2b is 
\[
\Pr\left[\ket{\psi}\to\ket{\psi}\right]=\Tr\left[\left(|\psi\rangle\langle\psi|\otimes\ii\right)|\Psi_{i}\rangle\langle\Psi_{i}|\right]
\]
 in which case the resulting state of the first register is $|\psi\rangle.$
The probability of measuring a zero in this step is 
\[
\Pr\left[\ket{\psi}\to\ket{\psi^{\perp}}\right]=\Tr\left[\left(|\psi^{\perp}\rangle\langle\psi^{\perp}|\otimes\ii\right)|\Psi_{i}\rangle\langle\Psi_{i}|\right]=1-\Pr\left[\ket{\psi}\to\ket{\psi}\right]
\]
 in which case the resulting state of the first register is $|\psi^{\perp}\rangle.$
Similarly, one can compute the transition probabilities starting from
the state $|\psi^{\perp}\rangle$ of the first register. These satisfy
\begin{align*}
\Pr\left[\ket{\psi^{\perp}}\to\ket{\psi^{\perp}}\right] & =\Pr\left[\ket{\psi}\to\ket{\psi}\right]\\
\Pr\left[\ket{\psi^{\perp}}\to\ket{\psi}\right] & =\Pr\left[\ket{\psi}\to\ket{\psi^{\perp}}\right]
\end{align*}
 so the random walk is symmetric. We can then directly apply our analysis
of the previous section to show that
\[
\mathbb{E}[\#\text{ of uses of phase estimation followed by measurement of \ensuremath{P}}]\leq2r.
\]

Each time we use phase estimation with $t$ ancillae, we use the gate
$W_{i}$ less than $2^{t}$ times \cite{NCbook}. So each time we
repeat phase estimation followed by measurement of $P$ we use less
than $2^{t}+1$ measurements of $P$ so the expected total number
of times $\mathbb{E}[M_{\text{PE}}^{(i)}]$ (PE stands for phase estimation)
that we use the measurement of $P$ is 
\begin{align*}
\mathbb{E}[M_{\text{PE}}^{(i)}] & <2r\cdot\left(2^{t}+1\right)\\
 & \le2r\left(\frac{12\pi}{\delta}+1\right)\\
 & =2\left\lceil \frac{1}{\log_{2}\left(\frac{2}{\sqrt{3}}\right)}\log_{2}\left(\frac{d}{2\epsilon}\right)\right\rceil \left(\frac{12\pi}{\delta}+1\right).
\end{align*}

Repeating this procedure to obtain estimates of each $q_{i}$ takes
expected running time 
\begin{equation}
\mathbb{E}[M_{\text{PE}}]<2d\left\lceil \frac{\log_{2}\left(\frac{d}{2\epsilon}\right)}{\log_{2}\left(\frac{2}{\sqrt{3}}\right)}\right\rceil \left(\frac{12\pi}{\delta}+1\right).\label{eq:numberPE}
\end{equation}

We now show that the probability that all the estimates $q_{i}^{est}$
obtained by using the above algorithm satisfy 
\[
|q_{i}^{est}-q_{i}|<\delta
\]
 is at least $1-\epsilon$. Our choice of $t$ was designed so that
the output of phase estimation of $W_{i}$ on the state $|\phi_{i}^{+}\rangle$
using $t$ ancillae is a state $|\phi_{i}^{+}\rangle|\gamma\rangle$
such that a measurement of the $t$-qubit state $|\gamma\rangle$
in the computational basis produces a phase $\tilde{\phi}$ that satisfies
\[
|\tilde{\phi}-\phi_{i}|\leq\frac{\delta}{3\pi}
\]
 with probability at least $\frac{3}{4}$\cite{NCbook}. Similarly
the output of phase estimation of $W_{i}$ on the state $|\phi_{i}^{-}\rangle$
using $t$ ancillae is a state $|\phi_{i}^{-}\rangle|\mu\rangle$
such that a measurement of the $t$-qubit state $|\mu\rangle$ in
the computational basis produces a phase $\tilde{\phi}$ that satisfies
\[
|\tilde{\phi}-(1-\phi_{i})|\leq\frac{\delta}{3\pi}
\]
 with probability at least $\frac{3}{4}$. In step 2(a) of our algorithm
we perform phase estimation on either the state $|\psi\rangle$ or
the state $|\psi_{i}^{\perp}\rangle$. In either case, the reduced
density matrix of the $t$-qubit ancilla register after applying the
phase estimation (but before measuring the phase) is 
\[
\frac{1}{2}\left(|\gamma\rangle\langle\gamma|+|\mu\rangle\langle\mu|\right)
\]
 which is an equal probabilistic mixture of $|\gamma\rangle$ and
$|\mu\rangle$. So, with probability at least $\frac{3}{4}$ (regardless
of whether we started in $|\psi\rangle$ or $|\psi^{\perp}\rangle$),
the phases $\phi_{i}^{(j)}$ measured in step 2 of the algorithm satisfy
either 
\[
|\phi_{i}^{(j)}-\phi_{i}|\leq\frac{\delta}{3\pi}
\]
 or 
\[
|\phi_{i}^{(j)}-(1-\phi_{i})|\leq\frac{\delta}{3\pi}.
\]
 Using the inequality 
\[
|\cos^{2}(\pi\alpha)-\cos^{2}(\pi\beta)|\leq2\pi|\alpha-\beta|\text{ }
\]
 and the fact that $\cos^{2}(\pi x)=\cos^{2}(\pi(1-x))$ it follows
that the estimates $q_{i}^{(j)}$ each (independently) satisfy 
\[
|q_{i}^{(j)}-q_{i}|<\delta
\]
 with probability at least $\frac{3}{4}$. The median lemma of \cite{NWZqma}
says in this case that the probability that the median of the $r$
independent measured values $q_{i}^{(j)}$ falls outside the interval
$(q_{i}-\delta,q_{i}+\delta)$ is upper bounded as $p_{\text{fail}}\leq\frac{1}{2}\left(\frac{\sqrt{3}}{2}\right)^{r}$
. Plugging in our choice of $r$ from Eq. \ref{eq:r_value} gives
\[
|q_{i}^{\text{est}}-q_{i}|<\delta
\]
 for each $i$ with probability at least $1-\frac{\epsilon}{d}$.
So the probability that the above inequality is satisfied for all
of the $i\in\{1,...d\}$ is at least $1-\epsilon$.

\subsection{Performance comparison for estimating measurement statistics}

These three algorithms for estimating the probabilities $q_{i}=\Tr\left[\rho_{B}\kets i_{B}\bras i_{B}\right]$
give estimates $\left\{ q_{i}^{\text{est}}\right\} $ (for $i$ from
$1$ to $d$) which are all within $\delta$ of the correct values
with probability at least $1-\epsilon$. Their running times are summarized
in table \ref{tab:Comparison-of-algorithms}.

\begin{table}
\begin{centering}
\begin{tabular}{|c|c|c|}
\hline 
State Restoration  & Alternating Projectors  & Phase Estimation\tabularnewline
\hline 
\hline 
\noalign{\vskip\doublerulesep}
$\mathbb{E}[M_{\text{SR}}]=\mbox{O\ensuremath{\left(\frac{\chi\cdot d}{\delta^{2}}\log\frac{d}{\epsilon}\right)}}$  & $\mathbb{E}[M_{\text{AP}}]=O\left(\frac{d}{\delta^{2}}\log\frac{d}{\epsilon}\right)$ & $\mathbb{E}[M_{\text{PE}}]=O\left(\frac{d}{\delta}\log\left(\frac{d}{\epsilon}\right)\right)$\tabularnewline[\doublerulesep]
\hline 
\noalign{\vskip\doublerulesep}
\end{tabular}
\par\end{centering}

\caption{Scaling of the expected number of measurements of $P=|\psi\rangle\langle\psi|$
used by each algorithm as a function of the desired precision $\delta$
and error probability $\epsilon$. \label{tab:Comparison-of-algorithms}}
\end{table}

State restoration is conceptually the simplest of the three algorithms,
and we expect that it will be sufficient for most purposes. It is
also the slowest as a function of $d$ and $\delta$ (assuming $\chi$
is increasing as a function of $d$). The state restoration algorithm
has the advantage that we can drop in different tomography schemes
that may improve performance.

In the absence of a better tomography scheme, however, both other
algorithms outperform state restoration as a function of $d$. Phase
estimation also performs quadratically better than both other algorithms
as $\delta\to0$.

\section{Applications of quantum state restoration and single-copy tomography}

\subsection{Breaking quantum money}

As we discussed in the introduction, quantum money is the idea of
using a state as money---that is, something that can be passed around
but not forged. The money consists of a quantum state and a verification
procedure which should succeed with high probability on valid money
issued by the bank but should fail with high probability for any efficiently
forgeable state. The first quantum money protocols \cite{BBBW83,Wie83}
required the verification procedure to be secret, so only the bank
(i.e.\ the issuer of the money) could verify money states. There
is recent interest in publicly verifiable quantum money \cite{aaronson-quantum-money,QMON09,mosca-2009},
in which everyone, including a would-be forger, has access to the
verification procedure. In the introduction, we showed that quantum
state restoration breaks quantum money based on product states. More
generally, as a corollary of Theorem~\ref{thm:single-copy-tomography},
any quantum money protocol in which the verifier is a projector must
be designed to withstand attacks based on single-copy tomography.
If the verifier is a projector, then an adversary can use single-copy
tomography to learn the measurement statistics of any efficiently
implementable measurement with a small number of outcomes on the quantum
money state $\ket{\psi}$.

\subsection{Studying ground states of many-body Hamiltonians}

Quantum computers offer potentially exponential speedups in simulating
quantum mechanics, but some problems are still hard. For example,
preparing ground states of many-body systems generically takes exponential
time in the number of particles. Nonetheless, for sufficiently small
systems with large enough energy gaps, algorithms such as \cite{poulin-wocjan-09}
may run quickly enough to prepare a single copy of the ground state,
and phase estimation can be used to verify the ground state. Single-copy
tomography allows us to make multiple tomographic measurements (even
of nonarxcommuting operators) on small numbers of particles without
having to prepare multiple copies of the ground state. This gives
a large speedup over traditional tomography.

Single-copy tomography could also be useful to characterize the ground
state during adiabatic evolution. This information could even be used
in real time to guide the choice of path for an adiabatic algorithm.

\section{Conclusions}

It is strongly believed that the ability to verify an unknown state
$\ket{\psi}$ does not give the ability to produce that state efficiently.
Without the ability to verify a state, mere possession of that state
confers little power. As we have shown, the combination of a verifier
and a single copy of $\ket{\psi}$ is more powerful that either one
alone.

\section{Acknowledgements}

This work was supported in part by funds provided by the U.S. Department
of Energy under cooperative research agreement DE-FG02-94ER40818,
the W. M. Keck Foundation Center for Extreme Quantum Information Theory,
the U.S. Army Research Laboratory's Army Research Office through grant
number W911NF-09-1-0438, the National Science Foundation through grant
number CCF-0829421, the NDSEG fellowship, the Natural Sciences and
Engineering Research Council of Canada, and Microsoft Research. D.N.\ gratefully
acknowledges support by European Project OP CE QUTE ITMS NFP 26240120009
and by the Slovak Research and Development Agency under the contract
No. APVV LPP-0430-09, and thanks Eddie Farhi's group for their hospitality.\bibliographystyle{hplain}
\bibliography{state_recovery}

\begin{thebibliography}{10}

\bibitem{aaronson-quantum-money}
S.~Aaronson.
\newblock Quantum copy-protection and quantum money.
\newblock In {\em Computational Complexity, Annual IEEE Conference on}, pages
  229--242, 2009.

\bibitem{QMON09}
Scott Aaronson, Edward Farhi, David Gosset, Avinatan Hassidim, Jon Kelner,
  Andrew Lutomirski, and Peter Shor.
\newblock Breaking and making quantum money: Toward a new quantum cryptographic
  protocol.
\newblock {\em Innovations in Computer Science ICS2010}, 0912.3825.

\bibitem{BBBW83}
C.H. Bennett, G.~Brassard, S.~Breidbart, and S.~Wiesner.
\newblock {Quantum cryptography, or unforgeable subway tokens}.
\newblock In {\em Advances in Cryptology--Proceedings of Crypto}, volume~82,
  pages 267--275, 1983.

\bibitem{Hoeffding}
Wassily Hoeffding.
\newblock Probability inequalities for sums of bounded random variables.
\newblock {\em Journal of the American Statistical Association}, 58(301), 1963.

\bibitem{JordansLemma}
C.~Jordan.
\newblock {\em Bulletin de la S. M. F.}, 3:103, 1875.

\bibitem{MWqma}
Chris Marriott and John Watrous.
\newblock {Quantum Arthur-Merlin games}.
\newblock {\em Computational Complexity}, 14(2):122--152, 2005.

\bibitem{mosca-2009}
Michele Mosca and Douglas Stebila.
\newblock Quantum coins, 2009.

\bibitem{NWZqma}
Daniel Nagaj, Pawel Wocjan, and Yong Zhang.
\newblock {Fast amplification of QMA}.
\newblock {\em Quantum Information \& Computation}, 9(11\&12):1053--1068, 2009.

\bibitem{NCbook}
Michael~A. Nielsen and Isaac~L. Chuang.
\newblock {\em Quantum Information and Computation}.
\newblock Cambridge University Press, Cambridge, UK, 2000.

\bibitem{poulin-wocjan-09}
David Poulin and Pawel Wocjan.
\newblock Preparing ground states of quantum many-body systems on a quantum
  computer.
\newblock {\em Physical Review Letters}, 102(13):130503, 2009, 0809.2705.

\bibitem{Wie83}
Stephen Wiesner.
\newblock Conjugate coding.
\newblock {\em SIGACT News}, 15(1):78--88, 1983.

\end{thebibliography}

\end{document}